\documentstyle[12pt,aasms]{article}
\input epsf
\lefthead{Rawson {\it et al.}}
\righthead{Cepheid distance to \n3621}
\begin{document}
\def \Vmod    {29.84\pm 0.18}
\def \Imod    {29.55\pm 0.18}
\def \muo     {29.13\pm 0.18}
\def \dist    {6.3\pm 0.7}
\def \EVI     {0.29\pm 0.03}
\def \lmc     {10.63\pm 0.09}
\def \lmcVu   {11.02\pm 0.08}
\def \lmcIu   {10.86\pm 0.07}
\def \VmodDo  {29.87\pm 0.18}
\def \ImodDo  {29.54\pm 0.18}
\def \muoDo   {29.06\pm 0.18}
\def \distDo  {6.1\pm 0.7}
\def \EVIDo   {0.33\pm 0.04}
\def \lmcDo   {10.56\pm 0.10}
\def \lmcVuDo {11.05\pm 0.08}
\def \lmcIuDo {10.85\pm 0.08}
\def \vi{{\it V--I\/}}
\def \bv{{\it B--V\/}}
\def \n3621{NGC$\,$3621}

\vspace*{-1 truein}

\title{The Extragalactic Distance Scale Key Project VIII.\\
The Discovery of Cepheids and a New Distance to\\
\n3621 Using the Hubble Space Telescope$^\dagger$}

\author{
Daya~M.~Rawson\altaffilmark{1},
Lucas~M.~Macri\altaffilmark{2},
Jeremy~R.~Mould\altaffilmark{1},
John~P.~Huchra\altaffilmark{2},
Wendy~L.~Freedman\altaffilmark{3},
Robert~C.~Kennicutt\altaffilmark{4},
Laura~Ferrarese\altaffilmark{5},
Holland~C.~Ford\altaffilmark{6},
John~A.~Graham\altaffilmark{7},
Paul~Harding\altaffilmark{4},
Mingsheng~Han\altaffilmark{8},
Robert~J.~Hill\altaffilmark{9},
John~G.~Hoessel\altaffilmark{8},
Shaun~M.~G.~Hughes\altaffilmark{10},
Garth~D.~Illingworth\altaffilmark{11},
Barry~F.~Madore\altaffilmark{12},
Randy~L.~Phelps\altaffilmark{3},
Abhijit~Saha\altaffilmark{13},
Shoko~Sakai\altaffilmark{12},
Nancy~A.~Silbermann\altaffilmark{12},
\& Peter~B.~Stetson\altaffilmark{14}}

\altaffiltext{1}{Mount Stromlo and Siding Spring Observatories,
Institute of Advanced Studies, Australian National University, Weston Creek,
ACT 2611, Australia}

\altaffiltext{2}{Harvard Smithsonian Center for Astrophysics, Cambridge, MA
02138, USA}

\altaffiltext{3}{Carnegie Observatories, Pasadena CA 91101, USA}

\altaffiltext{4}{Steward Observatory, University of Arizona, Tucson AZ 85721,
USA}

\altaffiltext{5}{Department of Astronomy, California Institute of Technology,
Pasadena CA 91125, USA}

\altaffiltext{6}{John Hopkins University and Space Telescope Institute,
Baltimore MD 21218, USA}

\altaffiltext{7}{DTM, Carnegie Institution of Washington, Washington, DC 20015,
USA}

\altaffiltext{8}{University of Wisconsin, Madison, WI, 53706, USA}

\altaffiltext{9}{Lab. for Astronomy and Solar Physics, NASA GSFC, Greenbelt MD
20771, USA}

\altaffiltext{10}{Royal Greenwich Observatory, Cambridge CB3 0EZ, UK}

\altaffiltext{11}{Lick Observatory, University of California, Santa Cruz CA
95064, USA}

\altaffiltext{12}{IPAC, California Institute of Technology, Pasadena CA 91125,
USA}

\altaffiltext{13}{Space Telescope Science Institute, Baltimore MD 21218, USA}

\altaffiltext{14}{Dominion Astrophysical Observatory, Victoria, BC V8X 4M6,
Canada}

\renewcommand{\thefootnote}{\fnsymbol{footnote}}
\footnotetext[2]{Based on observations with the NASA/ESA
{\it Hubble Space Telescope}, obtained at the Space Telescope Science
Institute, operated by AURA, Inc. under NASA contract No. NAS5-26555.}

\begin{abstract}
We report on the discovery of Cepheids in the field spiral galaxy NGC$\,$3621,
based on observations made with the Wide Field and Planetary Camera 2 on board
the {\it Hubble Space Telescope} (HST). \n3621 is one of 18 galaxies observed
as a part of {\it The HST Key Project on the Extragalactic Distance Scale},
which aims to measure the Hubble constant to 10\% accuracy. Sixty-nine
Cepheids with periods in the range 9--60 days were observed over 12 epochs
using the F555W filter, and 4 epochs using the F814W filter. The HST F555W and
F814W data were transformed to the Johnson $V$ and Kron-Cousins $I$ magnitude
systems, respectively. Photometry was performed using two independent
packages, DAOPHOT~II/ALLFRAME and DoPHOT.  \vskip12pt Period-luminosity
relations in the $V$ and $I$ bands were constructed using 36 fairly isolated
Cepheids present in our set of 69 variables. Extinction-corrected distance
moduli relative to the LMC of $\lmc$~mag and $\lmcDo$~mag were obtained using
the ALLFRAME and DoPHOT data, respectively. True distance moduli of $\muo$~mag
and $\muoDo$~mag, corresponding to distances of $\dist$~Mpc and $\distDo$~Mpc,
were obtained by assuming values of $\mu_0=18.50 \pm 0.10$~mag and $E(V-I) =
0.13$~mag for the distance modulus and reddening of the LMC, respectively.
\end{abstract}


\newpage

\section{Introduction}

These observations of the spiral galaxy \n3621 are part of a Key Project for
the {\it Hubble Space Telescope} (HST). This project, known as {\it The HST
Key Project on the Extragalactic Distance Scale} (\cite{ken95}), aims to
measure the Hubble constant to an accuracy of 10\%. By obtaining distances
based on Cepheid variable stars to eighteen different galaxies, we will
provide a firm basis for the calibration of several secondary distance
indicators: the Tully-Fisher relation, surface brightness fluctuations, the
planetary nebula luminosity function, the globular cluster luminosity
function, type II supernova expanding photospheres, and type Ia supernova
standard candles (see \cite{jac92} for a review of these methods). The HST,
with its enhanced resolution over ground-based telescopes and its position in
orbit which removes many of the problems plaguing terrestrial observers, is
the ideal instrument for the stringent demands of this type of project. 
Observations of Cepheids in M100 using HST (\cite{fer96}) have already 
yielded Cepheid distances to galaxies in the Virgo Cluster, thus displaying 
the capabilities of the WFPC2 instrument in detecting Cepheids in galaxies 
well outside of the Local Group. In addition, distances to M81, M101 (\cite{kel96}), and 
NGC925 (\cite{sil95}) have also been determined as a part of the Key Project.

\n3621 is a relatively isolated spiral with a morphological 
classification of Sc II.8 (\cite{RSA}) or Sc III-IV (\cite{RC3}) and a low
galactocentric redshift of $526$ km/s (\cite{RC3}), indicative of a
comparatively small distance relative to that of the Virgo cluster
(\cite{mou95}). Its complex pattern of partially resolved, irregular spiral
arms makes it an excellent candidate for the detection of Cepheids, while its
high inclination ($i=51^\circ$) and well-ordered H-I rotation (\cite {gar77})
indicates that it is an ideal object for the calibration of the Tully-Fisher
relation as applied to field spirals.  \vfill\pagebreak\newpage We describe
the observations and preliminary reductions in \S 2.  The photometry and
calibration of the instrumental magnitudes is discussed in \S 3. The search
for Cepheids using two independent photometric algorithms and the resulting
set of variables are described in \S 4. Period-luminosity relations and
distance moduli are presented and discussed in \S 5, and our conclusions are
given in \S 6.

\section{Observations and Reductions}

\subsection{Observations}

Observations of \n3621 using the Wide Field and Planetary Camera 2 (WFPC2)
system on the HST commenced on December 27, 1994 with the first of 24 F555W
(approximately Johnson V) images. The observations were split over 12 epochs
within a 60 day window; 11 of them were cosmic-ray split and one was a single
exposure. In addition, 9 F814W (approximately Kron-Cousins I) images divided
among four epochs, and 4 F439W images over two epochs were obtained. Given the
short total integration time and the very limited phase coverage of the F439W
data, it was not included in this data analysis.  Single short exposures of
180 seconds were obtained in the F555W and F814W filters to provide a
linearity check on the magnitudes derived by the photometric routines (see \S
3.3 for details). The WFPC2 footprint for the observations, superimposed on a
wide-field image of the galaxy (kindly made available by Sandage \& Bedke
1985), is shown in Figure~1, while a mosaic of the WFPC2 F555W images can be
seen in Figure~2.

The WFPC2 includes four $800\times800$ CCD detectors. Three of these, the Wide
Field Cameras, have a pixel size of 0.1 arcseconds as projected on the sky,
for a total field of view of $1.3\times1.3$ arcminutes each. The remaining
CCD, the Planetary Camera, has a pixel size of 0.046 arcseconds as projected
on the sky, for a total field of view of $36\times36$ arcseconds. A more
detailed description of the WFPC2 instrument can be found in {\it The HST WFPC2
Instrument Handbook} (\cite{bur94}). All observations were made with the
telescope guiding in fine lock with a stability of approximately 3 mas. The
gain and readout noise were 7 e$^{-}$/DN and 7 e$^{-}$, respectively. The CCD
was operated at a temperature of --88$^\circ$ C for all observations.

Exposure times and dates for each observation are given in Table~1.  The
sampling strategy, as discussed by Freedman et al.~(1994a), has been designed
specifically for the Key Project with the purpose of maximising the
probability of detecting a Cepheid with a period in the chosen window of
10--60 days. It follows a power-law distribution in time which provides an
optimum sampling of the light curves of Cepheids in this period range and
reduces the risk of aliased detections. Figure~3 shows the detection
efficiency of our optimum sampling; high values of variance imply a higher
probability of detecting Cepheids with those periods while lower values of
variance indicate lower probabilities of detection. Note that the variance
turnover at 65 days (due to the width of the observing window) implies that
this is the longest period that can be reliably determined.

\subsection{Data Reductions}

The HST data have been calibrated using the pipeline processing at the Space
Telescope Institute (STScI). The full reduction procedure, given in Holtzman
et al.~(1995a), consists of: a correction of A/D errors, the subtraction of a
bias level for each chip, the subtraction of a superbias frame, the
subtraction of a dark frame, a correction for shutter shading, and a division
by a flat field. The names of the STScI reference files used for this
calibration are listed in the notes to Table 1. Furthermore, each of the
frames was corrected for vignetting and geometrical distortions in the WFPC2
optics (using files kindly provided by P.~Stetson and J.~Holtzman,
respectively). Lastly, the images were multiplied by four and converted to
integers, to reduce disk usage and allow image compression. This conversion
led to an effective readout noise of 4e$^-$ and a gain of 1.75e$^-$/DN. A more
complete description of the calibration steps can be found in Hill et
al.~(1997) and Stetson et al.~(1997).

Holtzman et al.~(1995b) have described the effect on photometry of charge
transfer inefficiency (CTI) in the Loral WFPC2 chips. The principal effect at
very low light level is a loss of sensitivity from bottom to top of each chip
amounting to 0.04 mag for a star at row 800 relative to an equivalent star at
hypothetical row 0. Subsequent investigations (Holtzman, private
communication) have made use of the pre-flash capability built into WFPC2 (but
not a user commandable facility). Pre-flash intensities from 30 e$^-$ to 1000
e$^-$ were added to observations of standard stars in $\omega$ Centauri, and
aperture photometry was carried out systematically as a function of pre-flash
level and y-coordinate. This demonstrated that data with 0 e$^-$ background
show a 0.04 mag ramp relative to 30 e$^-$ (and brighter) background
levels. For our purposes, this confines the CTI problem to the calibration
results of Holtzman et al. and implies that no ramp correction is required for
any of our observations, where the background level is always greater than
70~e$^-$. Similarly, the observations of NGC$\,$2419 and Pal 4 used in the
determination of the ALLFRAME PSF (see below) have high background levels and
are not affected by CTI.

Hill et al.~(1997) have described a difference in the photometric zero points
of the WFPC2 depending on the exposure time of the observations that are used
for their determination. For exposures lasting less than about sixty seconds,
the zero points are systematically higher by $\sim 0.05$~mag (on average) than
those of exposures lasting more than 1300 seconds. The effect seems to arise
from a loss of 2 e$^-$/pixel per exposure (Casertano \& Baggett 1997),
which significantly affects the
magnitudes of faint stars on images with low background levels. Given the
length of our exposures and their high background levels, we use the
long-exposure zero points in this paper.

\section{Photometry and Calibration}

\subsection{Photometric Reductions}

Photometry of \n3621 was obtained by two groups implementing independent
software packages; D.M.R. and J.R.M. at Mount Stromlo Observatory, Australia
used DAOPHOT~II/ALLFRAME (\cite{ste94}), while L.M.M. and J.P.H.  at the
Harvard-Smithsonian Center for Astrophysics, Cambridge, USA used a version of
DoPHOT (\cite{sch93}) specially modified to deal with HST WFPC2 data
(\cite{sah96}). The reader is referred to those respective publications for a
detailed description of each photometry package.  Each package performs
photometry using independent techniques, and a comparison of results yields a
fundamental check on the validity of the software and the photometry. This
provides a powerful tool for detecting the presence of errors and biases which
might go undetected if only one package were used.

\subsubsection{DAOPHOT~II and ALLFRAME}

Both the DoPHOT and ALLFRAME software packages use profile-fitting procedures
to measure magnitudes.  DoPHOT constructs a PSF (Point Spread Function) using
the brightest stars in the frame being reduced, while DAOPHOT~II/ALLFRAME has
the ability to construct PSFs from bright stars in the frame being reduced or
from independent images of un-crowded fields. The PSFs used for the ALLFRAME
reductions were made using images of un-crowded regions of NGC$\,$2419 and
Pal 4 (\cite{hil96}), with the implicit assumption that any long-term
temporal variation of the PSF for a given filter/chip combination can be
absorbed into the aperture correction term (see below). Because of the very
high degree of crowding, with few, if any un-crowded stars in the NGC$\,$3621
frames, this is the preferred approach for the DAOPHOT~II/ALLFRAME reduction
track. These globular cluster PSFs have significant levels of background
light, and should be free of low light level CTI problems in the wings.

The ALLFRAME reduction followed the same steps described in previous papers
of this series (\cite {kel96}, \cite{fer96}, \cite{sil95}, \cite{hil96}), and
we give only a brief description here. The FIND routine in DAOPHOT~II is
initially applied to all of the images. This routine looks for stellar-like
objects in each image, and provides a filter against cosmic-rays, and
non-stellar objects. This filter basically screens against objects which
deviate from a circularly symmetric shape (their ``roundness''), and also
against objects whose profile deviates substantially from that of a gaussian,
by being too sharp, or flat (their ``sharpness''). These parameters can be
set to allow more or less flexibility in the filter. By imposing hard limits
in this first stage of the reduction we obtain the positions for about 1000
stars in each image, although many of these will still be cosmic rays which
have passed through the filter described above. These positions are used by
the DAOMASTER program to derive coordinate transformations between each of
the images. These are further refined using the ALLFRAME package on this
limited set of stars, which improves the accuracy of the
transformations. Using these transformations, a very deep image is obtained
by mosaicing all of the images together. FIND is then run on this image,
which should be free of contaminating cosmic rays. ALLSTAR is then used to
subtract each of these stars, and FIND is run again. This process is repeated
until no new stars are found. A final visual examination of the image reveals
a few remaining stars (typically around 30), which are also included in the
list. Using this final list of stars, ALLFRAME is used to obtain a final list
of photometric data. Using the epoch-epoch coordinates derived above, ALLFRAME
simultaneously fits a PSF to each star in each image, thus ensuring a
self-consistent determination of the magnitude across the window of
observations. The calculation of the sky background from an annulus around
each star is made only after {\it all} stars have been removed from the
image.

\subsubsection{DoPHOT}

A brief description of the DoPHOT reduction steps is given in this section;
Saha et al.~(1996) contains a detailed account of the process. The DoPHOT
reduction started by combining each pair of cosmic-ray split images to produce
a single, cosmic-ray-free image. This was done using a special algorithm
(developed by A. Saha and described in Saha~1996) that co-adds the cosmic-ray
split images and identifies pixels that exhibit a large variation between the
two frames. Special care is taken to bring both frames to the same background
level and to ensure that the program does not trigger on small, sub-pixel
shifts of the cores of bright stars. Once the contaminated pixels are
identified, they can be replaced with the corresponding pixels from the other
frame or set to a mask value so that they are not used at later stages of the
processing.

The version of DoPHOT optimised for HST photometry was developed by A. Saha to
address some aspects of WFPC2 data that are quite different from ground-based
CCD data, such as a point-spread function with a tight core and flared wings
and the removal of faint cosmic-ray hits and hot pixels, which is complicated
by the under-sampled nature of the data. A DoPHOT reduction starts with the
automatic identification of the brightest non-saturated stellar objects in the
frame, which are used to construct an analytic point-spread function (PSF).
DoPHOT then proceeds to examine every object whose central pixel is above a
certain threshold (which is lowered after each iteration). Every object in the
frame is analysed to determine if it is best fit by the stellar PSF, by two
close stellar PSFs (appropriate for blended stars), by an extended PSF
(appropriate for background galaxies or other extended objects), or by a
cosmic-ray profile. Once the best fit is determined, it is applied to the
object and used to remove it from the frame. At the end of an iteration,
the PSF is improved by using the information obtained from the newly identified
stars, and the program goes on to identify objects of lower threshold.

DoPHOT was run a first time on the cleaned images to identify bright objects,
which were used to derive coordinate transformations between the epochs.
The epochs with negligible shifts (i.e., less than one pixel in each
direction) were processed with the same algorithm described above to create
deep master frames. These master frames have several advantages, not only
in terms of signal-to-noise but also due to the fact that the co-addition
of several epochs dithers the sub-pixel profile of the PSF core into a
gaussian and allows DoPHOT to determine the centroid of an object with
greater accuracy. This reduces the uncertainty in the photometry, which
is greatly affected by centering uncertainties in the case of under-sampled
data.

The object lists obtained from the reduction of the master frames were
transformed into the coordinate systems of each epoch and used by DoPHOT to
obtain photometric information for each object from each frame.  The
magnitudes obtained by DoPHOT are ``fitted'' magnitudes, which need to be
tied to the standard 0.5-arcsecond magnitude system described in Holtzman
et al.~(1995b). This calibration is described next.

\subsection{Calibration}

The calibration of photometry for \n3621 was achieved using the method
outlined in Holtzman et al.~(1995b) and Hill et al.~(1997). The general
equations which convert the ``flight-system'' instrumental magnitudes into
the standard system (Johnson $V$ and Kron-Cousins $I$) are:
\vskip -18pt
\begin{eqnarray}
V&=&F555W-0.052(V-I)+0.027(V-I)^2 \\
I&=&F814W-0.063(V-I)+0.025(V-I)^2.
\end{eqnarray}

The preceding equations refer to instrumental magnitudes ($F555W$ and $F814W$)
that have been fully calibrated and corrected for exposure time and for
changes in the gain state of the camera. This is due to a change in the gain
state of the camera from 14~e$^-$/DN for the calibration data to 7~e$^-$/DN
for the \n3621 observations. The gain ratio terms (from \cite{hol95b}) are
1.987, 2.003, 2.006 and 1.955 for the PC1, WF2, WF3 and WF4 chips,
respectively, and have uncertainties of order 1\%.

The calibration of ALLFRAME photometry involves the corrections listed in the
preceding paragraph as well as an aperture correction to bring the fit
magnitudes to the 0.5-arcsecond magnitude system. This correction is obtained
by comparing photometric magnitudes for the brightest isolated stars in the
field of each chip within a 0.5-arcsecond radius calculated using DAOGROW
(\cite{ste90}) and the corresponding ALLFRAME magnitudes for those stars. The
number of stars used in this calculation were 22, 31, 32, and 27 in the PC1,
WF2, WF3, and WF4 chips, respectively. These aperture corrections are
averages over all our NGC 3621 frames. This local determination of the
aperture corrections allows one to correct the effect of any temporal
variation in the PSFs over the duration of cycle 4.  Variation in these
results from image to image is negligible, if the uncertainty in their
calculation is taken into account. Comparison with the results for other
galaxies in our cycle 4 series show a similar result with typical aperture
correction changes of 0.02 mag from galaxy to galaxy, changes which are
barely significant ($\sim2\sigma$) compared with their measuring errors.  The
final equations used to calibrate the ALLFRAME photometry are listed in
Table~A1.

The calibration of DoPHOT photometry also involves an aperture correction,
but this coefficient is calculated using a different approach (see Hill et
al.~1997 and Saha et al.~1996 for a complete description of the process).
It is known that the PSF of stellar cores in the WFPC2 chips varies slightly
as a function of position within the chip, while the aperture magnitudes
calculated using radii of five pixels or more are constant over the field of
view (\cite{hol95b}). Since DoPHOT uses a PSF that does not vary with
position, its fitted magnitudes will exhibit a variation of less than
0.15~mag peak to peak due to this effect. These variations are removed by
transforming the magnitudes into 5-pixel aperture magnitudes, using
correction coefficients derived by A.~Saha (\cite{sah96}) from observations
of the Leo I dwarf galaxy (\cite{mat96}). The transformation to
0.5-arcsecond aperture magnitudes is achieved using correction coefficients
derived from the Leo I data and the \n3621 master images. These coefficients
vary from frame to frame, and have values between 0.01~mag and 0.13~mag,
with uncertainties of $\sim 0.03$~mag.

\subsection{Comparison of ALLFRAME and DoPHOT Photometry}

A comparison of the ALLFRAME and DoPHOT photometric systems was conducted
for all chip/filter combinations using a visually-selected subset of stars
with no bright neighbours within five pixels. Their coordinates (obtained
with SAOIMAGE from the plate solutions provided by STScI in the image
headers) and $V$ and $I$ ALLFRAME and DoPHOT magnitudes are listed in Tables
A2-A5 for future comparisons of our photometry with other observations of
\n3621 Cepheids. Figure~4 shows plots of $\Delta$~mag (ALLFRAME-DoPHOT)
versus ALLFRAME magnitude for these stars. The mean offsets between the two
photometric systems are as follows: PC1, $\Delta V=-0.026\pm0.011$~mag and 
$\Delta I=-0.039\pm0.014$~mag; WF2, $\Delta V=+0.020\pm0.010$~mag and
$\Delta I=+0.071\pm0.010$~mag; WF3, $\Delta V=-0.066\pm0.011$~mag
and $\Delta I=+0.077\pm0.011$~mag; WF4, $\Delta V=-0.057\pm
0.010$~mag and $\Delta I=-0.018\pm0.010$~mag. This zero-point
difference has been seen in other galaxies observed as part of this Key
Project; it reflects uncertainties in the aperture corrections of different
filter/chip combinations as well as the different treatments of the sky
background by the two photometry packages. A full description of the
differences is outside the scope of the present paper; this will be analysed
in a forthcoming paper employing artificial stars to examine the systematics
of both photometry packages. Based on our results, we estimate a photometric
uncertainty of $\pm0.07$~mag in both bands for bright ($m < 24$), isolated
stars. See \S4.3 for a comparison of Cepheid magnitudes.

A linearity check was conducted using the 180-second $V$-band image taken
on JD 2449746. Fully-calibrated magnitudes of the brightest stars ($V < 23.5$
~mag) were compared with the mean magnitudes of the same stars. The zero-point
difference was found to be $\Delta V = 0.010\pm0.015$~mag and$\Delta V = 
0.066\pm0.094$~mag for ALLFRAME and DoPHOT respectively.

\section{Identification of Cepheid Candidates}

\subsection{Search for Cepheids}

The search for Cepheid variables using ALLFRAME and DoPHOT data was conducted
using similar but slightly different approaches. In both cases, the first
step was a classification of all objects as either variable or non-variable.
In the ALLFRAME search, this step involved a sigma-like classification of all
objects as either variable or non-variable, using a robust estimate of
$\sigma$, called $\sigma_F$ (\cite{hug89}; \cite{hoa83}), that is insensitive
to large outliers caused, for example, by cosmic rays. One obtains $\sigma_F$
by sorting all points in a light-curve in order of magnitude, and calculating
the spread (in magnitudes) of the middle 50\% of the sorted points. A star
was considered a variable if its $\sigma_F$ was greater than 2 times the mean
$\sigma_F$ of stars with similar magnitudes.  This method thus screens
against cosmic ray hits, producing a robust list of likely variables.

 The first step in the DoPHOT search was performed using a $\chi_\nu^2$
variability statistic,
\begin{equation}
\chi_\nu^2 = \sum^n_i \frac{(m_i - \bar{m})^2}{\nu\sigma^2_i},
\end{equation}
(\cite{sah90}), where $n$ is the number of data points available, $\bar{m}$ is
the mean magnitude of the star, $m_i$ and $\sigma_i$ are the individual
magnitude measurements and their uncertainties, and $\nu = (n-1)$. All objects
with with $\chi_\nu^2 \ge 1.75$ were flagged as as variable.

The second step in both searches consisted of a periodicity test of the
objects flagged as variables. The ALLFRAME search used a phase dispersion
minimisation (PDM) technique (\cite{ste78}), which is a variant on the Lafler
\& Kinman (1965) method.  The algorithm bins the light curve into multiple
phase and frequency bins, to compare the resultant dispersion with that
obtained for the un-binned light-curve.  For a periodic variable, this ratio
will then be a minimum at the relevant period and phase.  This technique is
ideal for detecting periodic, but minimally sampled, light-curves of any
shape.

The second step in the DoPHOT search used the original Lafler \& Kinman
method. A ``goodness of periodicity'' statistic $\Theta (P)$ is calculated,
\begin{equation}
\Theta (P) = \sum_i^n \frac{(m_{i+1} - m_i)^2}{(m_i - \bar{m})^2},
\end{equation}
where $P$ is a trial period, $\bar{m}$ is the mean magnitude of the star,
and the $m_i$'s are the individual magnitude measurements, ordered by phase
according to the trial period. $\Theta$ reaches a local minimum when $P$ equals
the period or an alias of the period; we chose the deepest minimum as the
variable period. Next, the $\Lambda$ parameter (defined as the ratio of
$\Theta$ away from the period to $\Theta$ at the period) was calculated;
only stars with $\Lambda \ge 3$ were considered {\it bona fide} periodic
variables.

The third and last step in both searches involved a visual examination of the
remaining objects to identify the Cepheid candidates. Period uncertainties
were estimated by visually inspecting the phasing of the light-curve for
nearby periods in the case of ALLFRAME, and the variation of $\Lambda$ with
trial period in the case of DoPHOT. To calibrate these estimated period
uncertainties, we adopted a Monte Carlo technique in which the light curves
were perturbed by the photometric uncertainties, and the periods re-derived.
The output period distributions of these simulations tend to be normal
distributions with sidelobes of widely varying amplitude due to
aliasing. Within the main peak of the distribution, the uncertainties quoted
in column (2) of Table 3 are approximately 2.5$\sigma$ error bars. But to
divide these quoted uncertainties by 2.5 to obtain $rms$ errors would be to
ignore the incidence of aliasing.

Both searches returned the same Cepheid candidates, with the exception of two
variables that are located in very crowded regions, and another located close to
the edge of the chip. For these three variables, only ALLFRAME photometry is
available. See \S~4.3 for details.

\subsection{Mean Magnitudes}

Phased light curves that are evenly sampled can be used to calculate mean
magnitudes by simply obtaining an intensity-averaged magnitude,
\vskip -18pt
\begin{equation}
m = -2.5 \log_{10}\frac{1}{N}\sum^{N}_{i=1}\;10^{-0.4\times m_i}.
\end{equation}

However, when the sampling is not uniform it is better to use a more elaborate
approach and obtain a phase-weighted magnitude,
\vskip -18pt
\begin{equation}
m = -2.5\log_{10}\sum^{N}_{i=1}\;0.5\times\left(\phi_{i+1}-\phi_{i-1}\right)
\times 10^{-0.4\times m_i},
\end{equation}
\noindent where $N$ is the total number of epochs and $m_i$ and $\phi_i$ 
are the magnitude and phase of the $i$-th epoch, in order of increasing
phase. Typical differences between the two methods are of the order 
a few hundredths of a magnitude, which do not impact significantly the
determination of the distance modulus.

In the case of cosmic-ray split pairs of observations, the two magnitude
measurements of a single epoch were combined using weights given by the
inverse square of the uncertainty in their photometry. This greatly reduces
contamination due to cosmic rays and spurious magnitude estimates. In
practice, obvious cosmic ray hits were removed from the data, their presence
being identifiable by a major increase in intensity for one of the cosmic-ray
split frames and a large uncertainty in the resulting photometry.

The $I$-band photometry consists of only four epochs of data, which gives rise
to under-sampling of the light curves and decreases the precision of both
methods of averaging. One way to compensate for this is to make use of the
correspondence between $V$ and $I$ light curves for Cepheids, as discussed by
Freedman (1988). This takes advantage of the fact that, as a first
approximation, one light curve can be mapped onto the other by simple
numerical scaling. The ratio of $V$ to $I$ amplitude is found to be 1:0.51.
Using this result, we can make an estimate of the correction to the average
$I$ magnitude that is required by the under-sampling. This is achieved by
calculating the difference between the $V$-band average for all epochs and for
just those epochs with both $V$- and $I$-band observations. This result is
rescaled by the 1:0.51 amplitude ratio and applied to the mean $I$ magnitude.
The correction is calculated using both methods of averaging and it amounts to
no more than $\pm0.1$ magnitudes, with an average of approximately $0.04$
magnitudes.

\subsection{The Cepheids Found in \n3621}

The ALLFRAME and DoPHOT variable searches produced a total of sixty-nine
Cepheids (identified as C01-C69). Table~2 lists, for each variable, its
identification number, the chip in which it is found, its J2000.0 coordinates
(obtained with SAOIMAGE from the plate solution provided by STScI in the
image headers) and comments on its quality. The comments describe the degree
of crowding (C, f, F, or FF), sampling of the light curve (A, a, or D) and
amplitude of the variability (B or E), and indicate the presence of other
irregularities such as high sky gradients (s), image defects (I) and cosmic
rays (d). A full key to these comments is given in the notes to the
table. Using this scheme, a Cepheid with high quality photometric data would
have the classification of ABC. This system was applied uniformly across the
entire sample of Cepheids to achieve a consistent determination of the best
variables to use in the final fit to the Period-Luminosity function.

Table~3 lists, for each variable, the ALLFRAME and DoPHOT periods and
phase-weighted magnitudes for the $V$ and $I$ bands. The individual
photometric measurements are listed in Tables~A6 and A7 for ALLFRAME $F555W$-
and $F814W$-band data, and in Tables~A8 and A9 for DoPHOT $F555W$- and
$F814W$-band data, respectively. No DoPHOT data is available for variable C07
(located close to one edge of one chip, in an area not searched by DoPHOT),
and for variables C34 and C46 (located within very crowded regions). DoPHOT
data is also unavailable for most variables in the epoch of JD 2449767 due to
problems with image registration.

A comparison of the ALLFRAME and DoPHOT $V$- and $I$-band phase-weighted
magnitudes for the Cepheids is presented in Figure~5. The mean offsets
(ALLFRAME - DoPHOT) are as follows: PC1, $\Delta
V=+0.07\pm0.02$~mag and $\Delta I=0.00\pm0.03$~mag; WF2, $\Delta
V=+0.01\pm0.03$~mag and $\Delta I=+0.09\pm0.03$~mag; WF3, $\Delta
V=-0.03\pm0.03$~mag and $\Delta I=0.14\pm0.05$~mag; WF4, $\Delta
V=-0.06\pm0.03$~mag and $\Delta I=0.07\pm0.03$~mag. While the
$V$-band differences are consistent with those derived from bright, isolated
field stars (\S3.3) except for the case of PC1 which is a three-sigma result. 
The $I$-band offsets also agree well within their uncertainties.  
These differences between the ALLFRAME and DoPHOT photometry sets are under 
investigation using experiments involving the
insertion of artificial stars in the data. It reflects no more than a 5\%
discrepancy in the distance, and is included in the error budget.

The Cepheids are identified in the field of each of the WFPC2 chips in
Figures~6a-d. Finding charts for each of the stars are displayed in
Figures~7a-l; each of the charts encompasses a 50 by 50 pixel area around
each variable. $V$- and $I$-band light curves based on the ALLFRAME data are
presented in Figures~8a-l. Figure~9 shows the location of all Cepheids in a
color-magnitude diagram of \n3621 stars. The Cepheids are shown to have
colors consistent with those expected for Cepheids.

\section{Period-Luminosity Relations and Distance Moduli}
\subsection{Methodology}
The method used to derive $V$- and $I$-band apparent distance moduli is the
same as that was used in other papers in this series. The period-luminosity
relations of LMC Cepheids (Madore \& Freedman 1991) are scaled to an assumed
true distance modulus of $\mu_{0,{\rm LMC}}$ = 18.50$\pm$0.10~mag (total
uncertainty) and corrected for an estimated average line-of-sight reddening of
$E(V-I)_{\rm LMC}=0.13$~mag (\cite{bes91}).

The period-luminosity relations can be expressed as:
\vskip -24pt
\begin{eqnarray}
M_V&=&-2.76\left(\log{\rm P}\left({\rm d}\right)-1.0\right)-4.16, \\
M_I&=&-3.06\left(\log{\rm P}\left({\rm d}\right)-1.0\right)-4.87.
\end{eqnarray}

In fitting the data from NGC$\,$3621, we follow the procedure given in 
Freedman et al.~(1994a). We fix the slope of best fit to that given in the 
equations above and scale the equations in the magnitude axis until the minimum
rms dispersion between the data and the fit is obtained. By using the slopes 
given above, we minimize any bias that would arise from incompleteness at faint
magnitudes in the \n3621 data set. We have also experimented with minimum period
cutoffs imposed on the data set following the discussion by Sandage etal. (1992). 
Variation in the distance modulus with period cutoffs up to 18 days is of the 
order of 0.04 mag.

The magnitude shifts resultant from this method yield $V$- and $I$-band
apparent distance moduli relative to the LMC.  A reddening correction due to
the presence of interstellar dust must be applied in order to obtain the
true distance modulus (see \cite{mad91} for a complete description of this
correction). The mean color excess $<E(V-I)>$ is effectively calculated by
returning each of the individual Cepheids to the ridge line of the intrinsic
period-color relation.  The total scatter in the individual estimates about
the average reddening is then effectively used to determine the quoted
uncertainty on the mean.  In reality however, that scatter contains also the
full intrinsic width of the period-color relation.  In determining the
reddening-corrected true modulus, the derived mean reddening is therefore
not used explicitly; rather the $V$ and $(V-I)$ data are combined in such a
way that a reddening-free magnitude\setcounter{footnote}{1}
\footnote{The factor of 2.45 can be simply derived from the extinction law 
of Cardelli et al. (1989). Using their notation we have
$A_I/A_V=0.7712-0.5897/R_V$, where $R_V=3.3$ for stars with colors similar
to those of Cepheids. A reddening free magnitude is thus defined by $W\equiv
V-A_V\equiv I-A_I$. Solving these two equations for $W$ yields the equation
above.} $W = V - 2.45(V-I)$ is calculated for each Cepheid and differenced
against the absolutely calibrated version of the $W-\log P$ relation. The
resulting true moduli are then averaged and a mean true modulus is
calculated.  It is the scatter about this quantity that defines the quoted
error in the true modulus.  This uncertainty can be significantly smaller
than the scaled error in the color excess because of the fact that lines of
constant period in the color-magnitude diagram are closely degenerate with
reddening trajectories.  That is, some fraction of the scatter in the
period-color relation (used to calculate reddenings and their errors) is
correlated with scatter in the period-luminosity relations and by removing
correlated errors due to reddening some (large) fraction of the
intrinsic-color-induced scatter is also removed.  The final scatter measured
in the individually derived true moduli primarily reflects errors in the
photometry plus the small degree to which the adopted reddening trajectory
departs from the intrinsic color-magnitude correlation.

\subsection{Results}
A subset of Cepheids with no bright stars within five pixels (i.e., class C
and f in Table~2) were selected for P-L fitting; there are 36 such stars with
ALLFRAME photometry, and one less in the DoPHOT set. Period-Luminosity
relations in the $V$ and $I$ bands are plotted in Figures~10 and 11 for
ALLFRAME and DoPHOT data, respectively. Filled circles are used to plot the
Cepheids in the selected subset, while the other ones are represented by open
circles.

The procedure described in \S5.1 yielded $V$- and $I$-band apparent distance
moduli relative to the LMC of $\lmcVu$~mag and $\lmcIu$~mag for the ALLFRAME
data, and $\lmcVuDo$~mag and $\lmcIuDo$~mag for the DoPHOT data. The quoted
uncertainties reflect {\it only} the rms dispersion of the \n3621
period-luminosity relations. The difference between ALLFRAME and DoPHOT 
$I$-band distance moduli is larger than the difference in the $V$-band
distance moduli, due to the larger discrepancy of $I$-band mean magnitudes
discussed in \S4.3.

Assuming a distance modulus of $18.50\pm0.10$~mag and a reddening of $E(V-I)=$
0.13~mag for the LMC, we obtained true distance moduli of $\muo$~mag and 
$\muoDo$~mag for the ALLFRAME and DoPHOT data, respectively, corresponding to  
distances of $\dist$~Mpc and $\distDo$~Mpc. These results are equivalent to  
$V$- and $I$-band apparent distance moduli of $\Vmod$~mag and $\Imod$~mag for 
the ALLFRAME data and $\VmodDo$~mag and $\ImodDo$~mag for the DoPHOT data, as 
well as mean reddening values of $E(V-I) = \EVI$~mag and $\EVIDo$~mag for the 
ALLFRAME and DoPHOT data, respectively.

The preceding results change slightly if the complete set of 69 (ALLFRAME) and
66 (DoPHOT) Cepheids is used: the ALLFRAME distance modulus increases by
0.07~mag, whereas the DoPHOT distance modulus remains the same. The resulting
distances are $6.5\pm0.7$~Mpc and $\distDo$~Mpc, respectively.

A full description of the different contributions to the uncertainties is
presented in Table~4. An important contribution to the error budget is the
uncertainty of the effect due to the different metallicities of the LMC and
\n3621 ([O/H] = --3.60 and --3.05, respectively). Assuming that the
coefficient $-\frac{\partial V}{\partial(\log Z)}|_P$ is less than
0.16~mag/dex (based on unpublished simulations by Wood, Mould and Madore),
we have estimated this uncertainty to be $\pm0.10$~mag.

For comparison with Equations (7) and (8), the \n3621 period-luminosity
relations based on ALLFRAME results are:
\vskip -24pt
\begin{eqnarray}
m_V&=&-2.76\left(\log{\rm P}\left({\rm d}\right)-1.0\right)+25.68\pm0.09\\
m_I&=&-3.06\left(\log{\rm P}\left({\rm d}\right)-1.0\right)+24.68\pm0.09.
\end{eqnarray}

\section{Conclusions}
We have discovered a total of 69 Cepheid stars in the Sc field spiral galaxy
NGC 3621. We have fitted standard $V$- and $I$-band period-luminosity
relations to a subset of the variables and obtained true distance moduli of
$\muo$ and $\muoDo$ magnitudes for the ALLFRAME and DoPHOT data,
respectively. These correspond to distances of $\dist$ and $\distDo$ Mpc. Mean
reddenings were deduced from the apparent $V$- and $I$-band moduli using a
standard reddening curve, and they amount to E(\vi) = $\EVI$ and $\EVIDo$
magnitudes. There is agreement, within the uncertainties, between the ALLFRAME
and DoPHOT results.

\section{Acknowledgements}
We would like to thank Doug Van Orsow, the program coordinator for this Key 
Project, as well as the rest of the STScI and NASA support staff that have made
this project possible. The HST Key Project on the Extragalactic Distance Scale 
is supported by NASA through grant GO-2227-87A from STScI. This work has made
use of the NASA/IPAC Extragalactic Database (NED), which is operated by the Jet
Propulsion Laboratory, Caltech, under contract with the National Aeronautics
and Space Administration.

\vfill\newpage\pagebreak
\begin{center}
{\bf Figure Captions}
\end{center}
\par
\noindent{Figure 1: -- A wide-field image of \n3621 (courtesy Sandage \& Bedke
1985), with the footprint of the WFPC\,2 field of view (see Figure~2)}\baselineskip=18pt

\noindent{Figure 2.-- A mosaic of the WFPC2 images of \n3621, showing the detailed 
structure of the region outlined in figure~1. Single-chip
images, marking the location of the variables, can be found in Figures~7a-d.}

\noindent{Figure 3.-- Sampling variance for the $V$-band observations. The
lower variance points indicate periods at which the phase coverage is less
uniform, and a corresponding decrease in the probability of detection. Due to 
the quadratic distribution in time of the observations, an enhanced performance 
in the detection of variables is obtained. The variance turns over due to the 
65 day observing window.}

\noindent{Figure 4.-- A comparison of the ALLFRAME and DoPHOT $V$ and $I$
magnitudes for bright and fairly isolated stars present in each chip. Filled
and open circles represent $V$ and $I$. The mean offsets 
between the two photometric packages for the four chips are as follows: PC1,
$\Delta V=-0.026\pm0.011$~mag and $\Delta I=-0.039\pm0.014$~mag;
WF2, $\Delta V=+0.020\pm0.010$~mag and $\Delta I=+0.071\pm0.010$~mag;
WF3, $\Delta V=-0.066\pm0.011$~mag and $\Delta I=+0.077\pm0.011$~mag; 
WF4, $\Delta V=-0.057\pm 0.010$~mag and $\Delta I=-0.018\pm
0.010$~mag. These differences are consistent with the results for
other galaxies examined in the Key Project.}

\noindent{Figure 5.-- A comparison of the ALLFRAME and DoPHOT $V$ and $I$
mean phase-weighted magnitudes for Cepheids in each chip. Filled and open
circles represent $V$ and $I$ magnitudes. The mean offsets and rms deviations
are as follows: PC1, $\Delta V=+0.07\pm0.02$~mag and $\Delta
I=0.00\pm0.03$~mag; WF2, $\Delta V=+0.01\pm0.03$~mag and $\Delta
I=+0.09\pm0.03$~mag; WF3, $\Delta V=-0.03\pm0.03$~mag and $\Delta
I=0.14\pm0.05$~mag; WF4, $\Delta V=-0.06\pm0.03$~mag and $\Delta
I=0.07\pm0.03$~mag.  The $V$- and $I$-band differences are comparable with
the results shown in Figure~4.}

\noindent{Figures 6a-d.-- $V$-band images of the four WFPC2 chips. The circles 
indicate the position of each of the Cepheids, labelled as in Table~2. Each
of the images is oriented such that the bottom left corner has the pixel 
coordinates 0,0 in each image.}

\noindent{Figures 7a-l.-- Finding charts for all the Cepheids. The position of 
each Cepheid is shown by the arrows. The coordinates are measured relative to
the bottom-left corner of the images in Figures 6a-d.}

\noindent{Figures 8a-l.-- $V$- and $I$-band phased light curves for all the
Cepheids, using ALLFRAME data. Filled and open circles represent $V$ and $I$
magnitudes, respectively. Two complete cycles are plotted to facilitate ease
of interpretation.}

\noindent{Figure 9.-- Color-magnitude diagram of stars in \n3621. Filled
circles are used to plot the subset of isolated Cepheids used to fit the 
period-luminosity relation, while open circles
represent the remaining Cepheids in the sample.}

\noindent{Figure 10.-- Period-luminosity relations in the V (top) and I (bottom) 
bands based on ALLFRAME photometry. Filled circles are used to plot the subset of 
isolated Cepheids with no companion star lying within a 5-pixel radius,
while open circles represent the crowded Cepheids. The solid lines are the best 
fits and the dotted lines correspond to the rms dispersion of the LMC 
period-luminosity relation of Madore \&~Freedman (1991). Apparent distance 
moduli obtained from the lines of best fit are $\Vmod$ and $\Imod$ in the 
V- and I-bands respectively. Including a correction due to reddening this corresponds 
to a distance of $\dist$ Mpc.}

\noindent{Figure 11.-- Period-luminosity relations in the V (top) and I (bottom) 
bands based on DoPHOT photometry. Filled circles are used to plot the subset 
of isolated Cepheids with no companion star lying within a 5-pixel radius, 
while open
circles represent the crowded Cepheids. The solid lines are the best fits and 
the dotted lines correspond to the rms dispersion of the LMC period-luminosity 
relation of Madore \&~Freedman (1991). Apparent distance moduli obtained from 
the lines of best fit are $\VmodDo$ and $\ImodDo$ in the V- and I-bands 
respectively. Including a correction due to reddening this corresponds 
to a distance of $\distDo$ Mpc.}

\vfill\pagebreak\newpage
\centerline{\bf Table Captions}

\noindent{Table 1: -- Log of HST WFPC2 observations of \n3621. All observations
comprise two cosmic-ray split pairs except for the epoch on January 4th. The
Julian Date quoted is the heliocentric Julian Date at mid-exposure.}

\noindent{Table 2: -- The Cepheid candidates found in \n3621. The columns 
provide the following information: (1) the identification; (2) the chip in 
which each variable is located; (3) and (4) the right ascension and 
declination in (J2000.0) and (5) the description of each of the Cepheids 
using the key given in the notes to the Table.}

\noindent{Table 3: -- The Cepheid candidates found in \n3621. The columns 
provide the following information: (1) the identification; (2) the ALLFRAME 
period in days; (3) and (4) the ALLFRAME $V$- and $I$-band phase-weighted 
magnitudes and their uncertainties (5) the; (DoPHOT period in days; (6) and 
(7) the DoPHOT $V$- and $I$-band phase-weighted magnitudes and their uncertainties.}

\noindent{Table 4: -- The contributions to the error budget of the 
distance modulus of \n3621.  The uncertainties in the $V$-band and $I$-band
P-L relation, are partially correlated ($\S$5.1) and this is taken into
account in calculating the uncertainty of the un-reddened P-L relation (R2).
This is in turn combined with the $rms$ error attributed to the photometric
algorithms to arrive at the uncertainty in the absolute distance modulus
relative to the LMC. This uncertainty is combined with the uncertainty in the
LMC distance modulus and the metallicity effect to arrive at the final
uncertainty in the distance modulus.}
\vskip 12pt
\centerline{\bf Appendix Table Captions}

\noindent{Table A1: -- The aperture correction and WFPC2 zero-point coefficients
used for ALLFRAME photometry.}

\noindent{Tables A2-A5: -- Secondary standard photometry for Chips 1-4}

\noindent{Table A6: -- $V$-band ALLFRAME data for each of the Cepheids over 
the twelve epochs.}

\noindent{Table A7: -- $I$-band ALLFRAME data for each of the Cepheids over 
the four epochs.}

\noindent{Table A8: -- $V$-band DoPHOT data for each of the Cepheids over 
the twelve epochs.}

\noindent{Table A9: -- $I$-band DoPHOT data for each of the Cepheids over 
the four epochs.}
\end{document}



\clearpage
\begin{deluxetable}{lcl}
\tablecolumns{3}
\tablewidth{15cm}
\tablenum{4}
\tablecaption{Error Budget for NGC 3621} 
\tablehead{
\colhead{Source of Uncertainty} & \colhead{Error} & \colhead{Comment}}
\startdata
{\bf CEPHEID PL CALIBRATION} & &  \nl
(a) LMC True Modulus & $\pm$ 0.10 & \nl
(b) V PL Zero Point & $\pm$ 0.05 & \nl
(c) I PL Zero Point & $\pm$ 0.03 & \nl
(S1) Systematic Uncertainty&$\pm$ 0.12&(a),(b),(c) combined in quadrature\nl
 & & \nl
{\bf NGC 3621 MODULUS} & &  \nl
(d) HST V-Band Zero Point & $\pm$ 0.05 & \nl    
(e) HST I-Band Zero Point & $\pm$ 0.05 & \nl   
(R1) Cepheid True Modulus & $\pm$ 0.15 &(d),(e) coupled by reddening law \nl
 & & \nl
(f) PL Fit (V) & $\pm$ 0.08 & \nl
(g) PL Fit (I) & $\pm$ 0.08 & \nl
(R2) Cepheid True Modulus &$\pm$ 0.10 &(f),(g) partially correlated \nl
 & & \nl
(S2) Metallicity Uncertainty&$\pm$ 0.10 & from M31 metallicity gradient test \nl
 & & \nl
{\bf TOTAL UNCERTAINTY} & &  \nl
(R) Random Errors&$\pm$ 0.18 &(R1) and (R2) combined in quadrature \nl
(S) Systematic Errors&$\pm$ 0.16 &(S1) and (S2) combined in quadrature \nl
\enddata
\end{deluxetable}

\begin{deluxetable}{ccccc}
\tablecolumns{5}
\tablenum{A1}
\tablewidth{11cm}
\tablecaption{Coefficients used for ALLFRAME Photometry}
\tablehead{
\colhead{Chip} & \colhead{Filter}  & \colhead{Aperture}   & \colhead{WFPC2}
& \colhead{C}\\
\colhead{}     & \colhead{}        & \colhead{Correction} & \colhead{Zeropoint}
& \colhead{Coeff.}}
\startdata
 PC1 & V & --\,0.067$\pm$0.02 & 22.510 & --1.036\nl
 PC1 & I & --\,0.077$\pm$0.03 & 21.616 & --1.940\nl
 WF2 & V & --\,0.014$\pm$0.02 & 22.522 & --0.971\nl
 WF2 & I &    +0.026$\pm$0.02 & 21.657 & --1.796\nl
 WF3 & V & --\,0.007$\pm$0.02 & 22.530 & --0.956\nl
 WF3 & I &    +0.032$\pm$0.03 & 21.638 & --1.809\nl
 WF4 & V & --\,0.011$\pm$0.02 & 22.506 & --0.984\nl
 WF4 & I &    +0.045$\pm$0.03 & 21.609 & --1.895\nl
\enddata
\tablecomments{The ALLFRAME instrumental magnitudes ($F555W$ and $F814W$ of
Eqns.~1-2) are obtained from the fitted magnitudes after applying the
correction $m_{\rm inst} = m_{\rm fit} + 2.5 * \log t + C$. The
coefficient $C$ equals $-25.00 + 0.016 + 2.5 * \log 4\ + AC + ZP$,
where $t$ is the exposure time, -25.00 is the ALLFRAME zeropoint,
+0.016 accounts for a pixel area normalization, $2.5 \log 4$ accounts
for the image conversion to integer data described in \S 2.1, $AC$ is
the aperture correction and $ZP$ is the WFPC2 zeropoint.}
\end{deluxetable}%
